\documentclass[fleqn,twoside]{article}
\usepackage{espcrc2,bm,multirow}

\newcommand{\AmS}{{\protect\the\textfont2  A\kern-.1667em\lower.5ex\hbox{M}\kern-.125emS}}
\newcommand{\tr}{\mathrm{tr}_\mathrm{CD}\!\ }
\newcommand{\SLASH}[1]{\slash\!\!\! #1}

\title{Flavourful hadronic physics}

\author{B.~El-Bennich\address[ANL]{Physics Division, Argonne National Laboratory, Argonne, Illinois 60439, USA},
              M.~A.~Ivanov\address{Bogoliubov Laboratory of Theoretical Physics, JINR, 141980 Dubna, Russia}
             and C.~D.~Roberts\addressmark[ANL]\address{Department of Physics, Peking University, Beijing 100871, China} }
        
\begin{document}

\begin{abstract}
We review theoretical approaches to form factors that arise in heavy-meson decays and are hadronic expressions of non-perturbative QCD. 
After motivating their origin in QCD factorisation, we retrace their evolution from quark-model calculations to non-perturbative QCD techniques 
with an emphasis on formulations of truncated heavy-light amplitudes based upon Dyson-Schwinger equations. We compare model predictions 
exemplarily for the $F^{B\to\pi}(q^2)$ transition form factor and discuss new results for the $g_{D^*\!D\pi}$ coupling in the hadronic $D^*$ decay.
\end{abstract}

\maketitle

\section{Flavour physics and strong phases}

The last two decades have witnessed important advances in flavour physics and in particular heavy-meson decays. From the first observation of a 
$B$ meson by the CLEO  Collaboration in 1981 at the Cornell Electron Storage Ring~\cite{Bebek:1980zd} (and their ongoing $D$-meson 
research program) to the dedicated $B$-physics facilities at SLAC in California and KEK in Japan, much progress has been made. Of course, 
while $B$ physics is the main focus of the Collaborations Belle at KEK and BaBar at SLAC, and of the CDF experiment at Fermilab, considerable 
efforts have also been devoted to studies of $D$-meson decays, charmonium and $\tau$ physics. 

Naturally, the driving force is to confirm the electroweak sector of the Standard Model which has established itself as the foremost paradigm
to describe $CP$ violation; in other words, the main task experimentalists strive for is the precise area of the Cabibbo-Kobayashi-Maskawa (CKM)
triangle and the weak $CP$ violating phase codified within its angles. In order to measure the size and exact form of this triangle, the angles are
derived from branching fractions in a nowadays large variety of decay channels. We focus on non-leptonic decays, $B\to M_1 M_2$, but our 
discussion also applies to the cleaner semi-leptonic $B\to M \ell\bar\nu_\ell$ case.

From a theoretical point of view, heavy mesons can be used to test simultaneously all the manifestations of the Standard Model, namely the interplay
between electroweak and strong interactions. They also provide an excellent playground to examine non-perturbative QCD effects already much 
studied in hadronic physics. It is noteworthy to remind that no $CP$-violating amplitude can be generated without strong phases. Suppose a heavy 
particle $H$ decays into a mesonic final state $M=M_1M_2 ...$, $H\to M$, and that the Standard Model lagrangian contributes two terms (two Feynman 
diagrams) to this process. Then, the decay amplitude and its corresponding $CP$ conjugate, written most generally, are
\begin{eqnarray}
  \mathcal{A}(H\to M) & = & \lambda_1 A_1 e^{i\varphi_1} + \lambda_2 A_2 e^{i\varphi_2} 
  \label{eq1}\\
   \bar{\mathcal{A}}(\bar H \to \bar M) & = & \lambda_1^* A_1 e^{i\varphi_1} + \lambda_2^* A_2 e^{i\varphi_2} \ .
  \label{eq2}
\end{eqnarray}
The weak coupling $\lambda_i$ is a combination of possibly complex CKM matrix elements and $A e^{i\varphi}$ denotes the strong (hadronic) parts 
of  the transition amplitude, where we emphasise that they too can have both a real part, or magnitude, and a phase, or absorptive part, due to multiple 
rescattering of the final-state quarks and mesons. These $CP$-related intermediate states must contribute the same absorptive part to the two decays, 
therefore the strong phases $\varphi_i$ are the same in Eqs.~(\ref{eq1}) and (\ref{eq2}). Taking the difference of the absolute squares, known as
{\em direct} $CP$ violation,
\begin{equation}
 |\mathcal{A}|^2  -  |\bar{\mathcal{A}}|^2   =   2 A_1 A_2\, \mathrm{Im} (\lambda_1 \lambda_2^*)  \sin (\varphi_1-\varphi_2 ) \ ,
\end{equation} 
one sees that {\bf no} such violation, $ |\bar{\mathcal{A}} /\mathcal{A} | \neq 1$, can occur if  the weak couplings contain only real phases or the 
strong phases are the same. Hence, in order to extract the weak CKM phases with precision from the decay amplitudes, it is crucial to evaluate 
the QCD contributions {\em reliably\/}.

\section{QCD factorisation}

As simple as these mesons appear to be---a bound colourless heavy-light $\bar qq$ pair---the difference in quark masses and the energetic light mesons
produced in their decays lead to a disparate array of energy scales. A central aspect of heavy-meson phenomenology are factorisation theorems which allow 
for a disentanglement of short-distance or {\em hard\/} physics, which encompasses electroweak interactions and perturbative QCD, from long-distance 
or {\em soft\/} physics, governed by non-perturbative hadronic effects. In the following, we illustrate the factorisation with non-leptonic decays of a heavy 
meson $H$.

In the hamiltonian formulation of heavy-quark effective theory (HQET)~\cite{Buchalla:1995vs}, in which amplitudes are expanded in powers of 
$\Lambda_\mathrm{QCD}/m_h$ and the heavy quark is a static particle in the limit $m_h\to \infty$, the $H\to M_1 M_2$ decay 
amplitude is given by
\begin{equation}
  \mathcal{A} = \frac{G_F}{\sqrt{2}}\sum_p \lambda_p\sum_i \,  C_i (\zeta) \langle M_1 M_2 | O_i | H \rangle (\zeta) \ ,
\end{equation}
where $\lambda_p =V_{pb}V^*_{pk}$  $(p=u,c; k=d,s)$ are products of CKM matrix elements and $G_F$ is the Fermi coupling constant. The dimension-six 
four-quark operators $O_i$ result from integrating out the weak gauge bosons $W^\pm$ in the operator product expansion and the Wilson coefficients 
$C_i(\zeta)$ encode perturbative QCD effects above the renormalisation point $\zeta$. 

In what is called ``naive" factorisation, the hadronic matrix element $\langle M_1 M_2 | O_i | H \rangle$ is approximated by the product of two bilinear currents, 
$\langle M_1| \bar q \gamma^\mu(1-\gamma_5) b | B\rangle$ $\otimes$ $\langle M_2 | \bar q' \gamma_\mu(1-\gamma_5) q | 0 \rangle$ +  $ (M_1 \leftrightarrow M_2)$,
where colour indices have been omitted. This factorisation simply expresses the matrix element of a local four-quark operator as a product of a decay 
constant and a transition form factor. However, as has long been known, the saturation by vacuum insertion fails in the case of most $D$ decay modes 
and is largely insufficient to reproduce the experimentally more precise data on $B\to M_1 M_2$ branching fractions. In fact, any hard final-state gluon 
interaction has been neglected and soft-gluon exchange is at best incorporated into an effective colour parameter or form factors. Moreover, the 
renormalisation scale and scheme dependence of $C_i(\zeta)$ are not cancelled by those of the matrix element $\langle M_1 M_2 | O_i | H \rangle (\zeta)$.

A major improvement over this simple factorisation {\em Ansatz\/} comes from the systematic reorganisation of weak and QCD interactions in the HQET. 
Three distinctive approaches have emerged in recent years: QCD factorisation (QCDF)~\cite{Beneke:1999br}, perturbative QCD (pQCD)~\cite{Keum:2000wi} 
and soft-collinear effective theory (SCET)~\cite{Bauer:2000yr}. We here focus on the form factors that emerge in these factorisation approaches
and solely remark that QCD corrections beyond naive factorisation entail, in the limit $m_h \gg \Lambda_\mathrm{QCD}$, the $B\to M_1 M_2$ decay 
amplitude can be schematically written as
\begin{eqnarray}
  \lefteqn{\langle M_1 M_2 | O_i | B \rangle\  = \   \langle M_1| j_1 | B \rangle \langle M_2 | j_2 | 0 \rangle }  \nonumber \\
    & &   \hspace*{5mm}  \times \ \left [ 1 + \sum_n r_n \alpha_s^n + \mathcal{O} ( \Lambda_\mathrm{QCD}/m_b ) \right ]  ,
\label{QCDfac}
\end{eqnarray}
where $j_1$ and $j_2$ are the bilinear currents. This has been shown explicitly to leading order in $\alpha_s$~\cite{Beneke:1999br,Bauer:2004tj}
and including the one-loop correction $(\alpha_s^2)$ to the tree-diagram scattering between the emitted meson and the one containing the 
spectator quark~\cite{Beneke:2005vv}. 

The factorisation theorem derived using SCET agrees with QCDF  if perturbation theory is applied at the hard $m^2_b$ and hard-collinear 
$m_b\Lambda$ scales, with $\Lambda$ typically of the order of 100~MeV. It is evident from Eq.~(\ref{QCDfac}) that higher orders in $\alpha_s$ 
break the factorisation but these corrections can be systematically supplemented; the analogy with perturbative factorisation for exclusive processes 
in QCD at large-momentum transfer is not accidental~\cite{Lepage:1980fj}. Further contributions that break the factorisation, formally suppressed in 
$\Lambda_\mathrm{QCD}/m_b$ yet not irrelevant, are weak annihilation decay amplitudes and final-state interactions between daughter 
hadrons~\cite{ElBennich:2009da}. Neglecting power corrections in $\alpha_s$ and taking the limit $m_b\to \infty$, the naive 
factorisation is recovered.

\section{Separating scales: the softer the harder}

While factorisation theorems elaborated with SCET provide the means to systematically integrate out energy scales in the perturbative domain, yielding
approximations valid in the heavy-quark limit for a given decay in terms of products of hard and soft matrix elements, a reliable evaluation of the latter 
is notoriously difficult. In fact, it is the soft physics of the bound states that renders the task hard, as it implies non-perturbative QCD contributions. 
Full {\em ab initio\/} calculations are currently out of reach and for the time being one is left with modelling the heavy-to-light amplitudes 
with as much input from non-perturbative QCD as possible. Just how much soft physics is included depends on the theoretical {\em Ansatz\/} and 
techniques employed.

A variety of theoretical approaches have been applied to this problem, recent amongst which are analyses using light-front and relativistic quark models, 
light-cone sum rules (LCSR) and lattice-QCD simulations. In Section~\ref{models} we briefly summarise these approaches while Section~\ref{DSE} deals 
in more detail with studies of heavy-to-light form factors within the framework of the Dyson-Schwinger equation (DSE). We refer to a recent 
review~\cite{Buchalla:2008jp} for a summary of transition form factor data from lattice-regularised QCD  and just note that current results 
are obtained at large squared momentum transfer, {\em i.e.\/}, $q^2 \simeq 16$~GeV$^2$ for $B\to \pi$ transitions. Hence, values at low $q^2$ must
necessarily be extrapolated by means of a (pole-dominance) parametrisation~\cite{Becirevic:1999kt,Ball:2004ye}.

\begin{table*}[t]
\vspace*{-1mm}
\begin{center}
\begin{tabular}{c|ccccccc}
$q^2$~[GeV$^2$]  &  Ref.~\cite{Ebert:2006nz} & Ref.~\cite{Melikhov:2001zv}&  Ref.~\cite{Lu:2007sg}  & Ref.~\cite{Ivanov:2007cw} 
 & Ref.~\cite{Khodjamirian:2006st}  & Ref.~\cite{Ball:2004ye} &   Ref.~\cite{DeFazio:2005dx} \\
\hline
$0$  & 0.217 & 0.29 & 0.247  &  0.24 &  $0.25\pm0.05$ & $ 0.26\pm 0.03$ & $0.27\pm 0.02\pm0.07$ \\
$10$ & 0.41 & 0.54  & 0.49    &  0.53  &  0.54  &  0.51  & --   \\
$15$ & 0.67 & 0.84  & 0.97    &  0.91  &  0.90  &  0.81  & --  \\
$20$ & 1.40 & 1.56  & $>10$&  1.75   & 1.83  &  1.58  & --   \\
\end{tabular}
\end{center}
\caption{\small Numerical comparison for the transition form factor, $F_+^{B\to\pi}(q^2)$; the $q^2$ values of Ref.~\cite{Ivanov:2007cw} are calculated,
whereas for Refs.~\cite{Ebert:2006nz,Melikhov:2001zv,Lu:2007sg,Khodjamirian:2006st,Ball:2004ye} the value $F_+^{B\to\pi}(0)$ and the corresponding
extrapolation in these references are employed.}
 \label{table1}
\vspace*{-1mm}
\end{table*}

\vspace*{-1mm}
\section{Hadronic transition form factors\label{models}}

\textit{Quark models\/}: 
relativistic quark models~\cite{Ebert:2006nz,Ivanov:2000aj,Faessler,Ivanov:2002un,Melikhov:2001zv,ElBennich:2008xy,Lu:2007sg} have in common 
that their only degrees of freedom are constituent quarks whose masses are parameters of the hamiltonian. The hadronisation of the two valence quarks 
is described by vertex wave functions or phenomenological Bethe-Salpeter amplitudes (BSA). The approaches 
in~\cite{Ivanov:2000aj,Faessler,Ivanov:2002un} represent heavy-to-light transition amplitudes by triangle diagrams,  a 3-point function between two 
meson BSA and the weak  coupling, which yields the transition amplitude $\langle M(p_2) | \bar q\, \Gamma_I h | H(p_1) \rangle$ and reads generally, 

\begin{eqnarray}
  \lefteqn{\mathcal{A}(p_1,p_2) = \tr \!\int\! \frac{d^4k}{(2\pi)^4} \, \bar \Gamma_{M}^{(\mu)}(k;-p_2)  S_q(k+p_2)} \nonumber   \\
  &  \times &   \Gamma_I(p_1,p_2)  S_Q (k+p_1)  \Gamma_H(k; p_1) S_{q'} (k) \ ,  
 \label{heavylightamp}
\end{eqnarray}
where $S(k)$ are quark propagators, $Q=c,b$; $q=q'=u,d,s$; $M=S,P,V,A$ and the index $\mu$ indicates a possible vector structure in the final-state BSA. 
$\Gamma_I$ is the interaction vertex whose Lorentz structure depends on the operator $O_i$ in the HQET and $\Gamma_H$ is the heavy meson BSA. 
The trace is over Dirac and colour indices. 

An analogous approach represents the amplitude in Eq.~(\ref{heavylightamp}) by relativistic double-dispersion 
integrals over the initial- and final-mass variables $p_1^2$ and $p_2^2$, where the integration kernel arises from the double discontinuity of the
triangle diagram (putting internal quark propagators on-shell via the Landau-Cutkosky rule). The meson-vertex functions are given by one-covariant 
BSA~\cite{Melikhov:2001zv,ElBennich:2008xy}. Other quark models~\cite{Ebert:2006nz} represent $B\to M$ form factors by overlap 
integrals of meson wave functions, obtained from confining potential models, and appropriate weak interaction vertices. Similar quark model calculations 
were performed on the light cone~\cite{Lu:2007sg}.

All these approaches model soft contributions with vertex functions, while the propagation of the constituent quark, $S(k)=(\SLASH k -m_q)^{-1}$, is 
scale independent and does {\em not\/} describe confinement and dynamical chiral symmetry breaking (DCSB). As noted in 
Refs.~\cite{ElBennich:2008xy,ElBennich:2008qa,Ivanov:2007cw}, this can lead to considerable model dependance at larger momentum transfer.
\smallskip

\textit{Light-cone sum rules\/}:  In a LCSR the operator-product expansion of a given correlation function is combined with hadronic dispersion relations. 
The quark-hadron duality is invoked: the correlator function is calculated twice, as a hadronic object and with subhadronic degrees of freedom. 
After separation in HQET of the heavy meson's static part, $P_H = p+q = m_h v_h+k$, where $v_h$ is the four-velocity and $k$ is the residual 
momentum, and likewise redefinition of the momentum transfer $q = m_h v_h + \tilde q \Rightarrow  p + \tilde q = k$, one obtains the heavy-limit
correlation function,
\begin{eqnarray}
\lefteqn{\Pi^H (p,q)  = \tilde \Pi^{H_v} (p,\tilde q) +  \mathcal{O} (1/m_h) ;}
\label{LCSR1} \\
\lefteqn{\tilde \Pi^{H_v} (p,\tilde q)  =  i\!\! \int\!  d^4\!x\, e^{ipx} \langle 0 | T [ J_M(x)J_{h_v}(0) ] | H_v  \rangle , } \nonumber
\end{eqnarray}
whereas a hadronic correlator can be written,
\begin{eqnarray}
\tilde \Pi^{\mathrm{had.}} (p,q) = \frac{\langle 0 | J_M | M(p) \rangle \langle M(p) | J_h | H (P_H) \rangle}{m_M^2 - p^2}  ,
\label{LCSR2}
\end{eqnarray}
where $J_h(0)$ and  $J_{h_v}(0)$ are heavy-light currents and $J_M(x)$ the interpolating current for a pseudoscalar or vector meson. 

In Eq.~(\ref{LCSR2}) only the light-meson contribution is represented but higher and continuum states can also be taken into account. In Eqs.~(\ref{LCSR1}) 
and (\ref{LCSR2}), the usual role of the correlation functions has been reversed~\cite{DeFazio:2005dx,Khodjamirian:2006st}: the correlation function 
is taken between the vacuum and the on-shell $B$-state vector using its light-cone distribution amplitude (DA) expansion and the pion is interpolated 
with the light-quark (axialvector) current $J_M(x)$. The $B$-meson DAs are universal non-perturbative objects introduced within HQET. 

In Ref.~\cite{Ball:2004ye}, however, the  correlation function is taken between the vacuum and the light-meson state, whereas the $B$ meson is 
interpolated by the heavy-light quark current $J_h(0)$.  As a result, the long-distance dynamics in the correlation function is described by a set of 
light-meson ($\pi, K, \rho, K^*$) DA. In the last step, a Borel transformations is applied to both, Eqs.~(\ref{LCSR1}) and (\ref{LCSR2}), from which 
one derives a transition form factor expressed as a sum rule. The transformation introduces a scale via the Borel parameter which, in turn, is fixed 
with sum rules for light-meson decay constants.

Besides a systematic uncertainty owing to the duality assumption, the main incertitude lies within the DAs which encode the {\em relevant\/} 
non-perturbative effects. Only their asymptotic form is known exactly from perturbative QCD. As of yet, the first two Gegenbauer moments of the DA 
for various light pseudoscalar and vector mesons have been obtained from QCD sum rules with very large errors, though the first moment is consistent 
with lattice calculations~\cite{Boyle2006}.  Moreover, the transition form factors must be extrapolated to space-like momenta.

For purpose of comparison, we list $B\to \pi$ transitions form factors, $F_+(q^2)$, for various models in Table~\ref{table1}.  As observed therein, 
there is a 30\% variation within the quark models~\cite{Ebert:2006nz,Melikhov:2001zv,Lu:2007sg} at $q^2=0$ which increases at larger $q^2$
values. The LCSR predictions~\cite{Khodjamirian:2006st,Ball:2004ye} agree at $q^2=0$ but their respective slopes for $q^2>0$ vary
by $12\%$. The form factors obtained with the DSE model~Ref.~\cite{Ivanov:2007cw}  are calculated on the entire physical momentum domain 
and the chiral limit is directly accessible.

\section{Flavourful Dyson-Schwinger equations\label{DSE}}

The elements entering the amplitude in Eq.~(\ref{heavylightamp}) can be motivated by the solutions of DSEs applied to QCD. A general review of the 
DSEs can be found in Refs.~\cite{Roberts:1994dr,Roberts:2007jh} and their applications to heavy-light transition form factors have been investigated 
in~\cite{Ivanov:2007cw,Ivanov:1997iu,Ivanov:1998ms}.
\smallskip

\textit{Dressed quark propagator\/}:
The mesons are bound states of a quark and antiquark pair, where for a given quark flavour their dressing is described by the DSE (in Euclidean metric),
\begin{equation}
S^{-1}(p)  =   Z_2 (i\gamma\cdot p + m^{\mathrm{bm}}) + \Sigma (p^2) \ ,
\label{DSEquark}
\end{equation}
with the dressed quark self energy,
\begin{equation}
 \Sigma (p^2) = Z_1 g^2\! \!\int^\Lambda_k \!\!\! D^{\mu\nu}Ê(p-k) \frac{\lambda^a}{2} \gamma_\mu S(k) \Gamma^a_\nu (k,p) ,
\end{equation}
where $\int_k^\Lambda$ represents a Poincar\'e invariant regularisation of the integral with the regularisation mass scale $\Lambda$.
The current quark bare mass $m^{\mathrm{bm}}$ receives corrections from the self energy $\Sigma (p^2)$ in which the integral is over the dressed 
gluon propagator, $D_{\mu\nu}(k)$, the dressed quark-gluon vertex, $\Gamma^a_\nu (k,p)$, and $\lambda^a$ are the usual SU$(3)$ colour 
matrices. The solution to the gap equation~(\ref{DSEquark}) reads\vspace*{-1mm}
\begin{eqnarray}
  S(p)&  = & -i \gamma\cdot p \ \sigma_V (p^2) + \sigma_S(p^2) \nonumber \\
  & = & \left [ i \gamma\cdot p \  A(p^2) + B(p^2) \right ]^{-1}  .
  \label{sigmaSV}
\end{eqnarray}

The renormalisation constants for the quark-gluon vertex, $Z_1(\zeta,\Lambda^2)$, and quark-wave function, $Z_2(\zeta,\Lambda^2)$, depend on
the renormalisation point, $\zeta$, the regularisation scale, $\Lambda$, and the gauge parameter, whereas the mass function $M(p^2) = B(p^2)/A(p^2)$ 
is independent of $\zeta$. Since QCD is asymptotically free, it is useful to impose at large spacelike $\zeta^2$ the  renormalisation condition,
\begin{equation}
  S^{-1}(p)|_{p^2=\zeta^2}  = i\gamma\cdot p + m(\zeta^2) \ ,
\end{equation}
where $m(\zeta^2)$ is the renormalised running quark mass, so that for $\zeta^2\gg \Lambda_\mathrm{QCD}^2$ quantitative matching with pQCD
results can be made.

Infrared dressing of light quarks has {\em profound\/} consequences for hadron phenomenology~\cite{Roberts:1994hh}: the quark-wave function renormalisation, 
$Z(p^2) = 1/A(p^2)$, is suppressed whereas the dressed quark-mass function, $M(p^2) = B(p^2)/A(p^2)$, is enhanced in the infrared which expresses 
dynamical chiral symmetry breaking (DCSB) and is crucial to the emergence of a constituent quark 
mass scale. Both, numerical solutions of the quark DSE and simulations of lattice-regularised QCD~\cite{Zhang:2004gv}, predict this behaviour of $M(p^2)$ 
and pointwise  agreement between DSE and lattice results has been explored in Refs.~\cite{Bhagwat:2003vw,Alkofer:2003jj}. Studies that do not 
implement light-quark dressing run into artefacts caused by rather {\em large} light-quark masses~\cite{ElBennich:2008xy,ElBennich:2008qa} to emulate 
confinement. This is because unphysical thresholds in transition amplitudes can only be overcome with the prescription that  $m_H < m_{q_1}+ m_{q_2}$,
which poses problems for a description of light vector mesons ($\rho, K^*$), heavy flavoured vector mesons ($D^*,B^*$) and for $P$-wave and excited 
charmonium states.

Whereas the impact of gluon dressing is striking for light quarks, its effect on the heavy quarks is barely notable. This can be appreciated, for instance, 
in Fig.~1 of Ref.~\cite{Ivanov:1998ms}: for light quarks, mass can be generated from nothing, {\em i.e.\/}, the Higgs mechanism is irrelevant to their acquiring 
of a constituent-like mass.
\smallskip

\textit{Bethe-Salpeter amplitudes\/}: The BSA can be determined reliably by solving the Bethe-Salpeter equation (BSE) in a truncation scheme consistent 
with that employed in the gap equation~(\ref{DSEquark}). Consider the inhomogeneous BSE for the axialvector vertex $\Gamma^{fg}_{5\mu}$ in which 
pseudoscalar and axialvector mesons appear as poles: \vspace*{-1mm}
\begin{eqnarray}
\label{BSE}
\lefteqn{ \Gamma^{fg}_{5\mu}(k;P) = Z_2 \gamma_5 \gamma_\mu - g^2\! \int_q^\Lambda\!\! D^{\alpha\beta}Ê(k-q) \frac{\lambda^a}{2} \gamma_\alpha }
 \nonumber \\
 \lefteqn{\times \ S_f(q_+) \,\Gamma^{fg}_{5\mu}(q;P) S_g(q_-) \frac{\lambda^a}{2} \Gamma^g_\beta (q_-,k_-) }\\
 \lefteqn{ +\   g^2\!   \int_q^\Lambda\!\!\! D^{\alpha\beta}Ê(k-q) \frac{\lambda^a}{2} \gamma_\alpha S_f(q_+) \frac{\lambda^a}{2} \Lambda^{fg}_{5\mu\beta}(k,q;P)\, .} 
 \nonumber 
\end{eqnarray}
In Eq.~(\ref{BSE}), $P$ is the total meson momentum, $q_\pm=q\pm P/2, k_\pm = k\pm P/2$, $\Lambda^{fg}_{5\mu\beta}$ is a 4-point Schwinger function 
entirely defined via the quark self energy~\cite{Chang:2009zb} and $f,g$ denote the flavour indices of a light-light or heavy-light bound state. The solutions 
of the vertex $\Gamma^{fg}_{5\mu}$ must satisfy the axial-vector Ward-Takahashi identity,
\begin{eqnarray}
  P^\mu\,  \Gamma^{fg}_{5\mu}(k;P)  =  S_f^{-1}(k_+)i\gamma_5 +i\gamma_5 S_g^{-1}(k_-) \nonumber \\
   - \   i [ m_f(\zeta) + m_g (\zeta )] \ \Gamma^{fg}_{5}(k;P) \  ,
\label{WTI}
\end{eqnarray}
where $\Gamma^{fg}_{5}$ solves the pseudoscalar analogue to Eq.~(\ref{BSE}). A systematic, symmetry-preserving truncation of the DSE and
BSE is given by the Rainbow ladder~\cite{Munczek:1994zz} which is their leading-order term with the dressed quark-gluon vertex,  $\Gamma^f_\mu$,
replaced by $\gamma_\mu$. It can be shown that $\Lambda^{fg}_{5\mu\beta} \equiv 0$ in this approximation.

The above Ward-Takahashi identity possesses another remarkable property; the set of quark-level Goldberger-Treiman relations that follow from it 
reveal the full structure of the light-meson BSA~\cite{Maris:1997hd}. In particular, it enables one to relate the leading covariant, $\varepsilon_5(k;P)$, of
the light pseudoscalar BSA with the scalar part, $B(p^2)$, of the dressed-quark propagator~(\ref{sigmaSV}) in the chiral limit. This motivates an effective
parametrisation of the light mesons ($M=\pi,K$),

\begin{equation}
 \Gamma_M (k;P) =  i \gamma_5\, \varepsilon_M(k^2)  =   i \gamma_5 \, B_M(k^2)/ \hat f_M  \ ,
 \label{BSApion}
\end{equation}
$\hat f_M= f_M/\sqrt{2}$, which has been capitalised on in transition form factor calculations~\cite{Ivanov:2007cw,Ivanov:1997iu,Ivanov:1998ms}.

Simultaneous solutions of the quarks's DSE and the heavy meson's BSA with renormalisation-group improved ladder truncations, obtained for the 
kaon~\cite{Maris:1997hd}, prove to be difficult. The truncations do not yield the Dirac equation when {\em one\/} of the quark masses is large. 
A recent attempt to calculate BSA for $D$ and $B$ mesons~\cite{Nguyen:2009if} reproduces well the respective masses but underestimates 
experimental leptonic decay constants by $30-50$\%. With a consistent derivation of the heavy meson BSA pending, simple one-covariant forms 
for $\Gamma_H (k;P)$ are currently employed in Eq.~(\ref{heavylightamp}), which reproduce leptonic decay constants in a simultaneous calculation. 

\section{Hadronic decays}

The decay $D^*\to D\pi$ can be used to extract the strong coupling $\hat g$ between heavy vector and pseudoscalar mesons to a low-momentum 
pion in the heavy meson chiral lagrangian~\cite{Wise:1992hn}. One considers the matrix element, 
\begin{equation}
  \langle  D(p_2) \pi( q) | D^*(p_1,\lambda) \rangle = g_{D^*\!D\pi} \  \bm{\epsilon}_\lambda\!\cdot q \ ,
\label{H*Hpi}
\end{equation}
where the coupling, $g_{D^*\!D\pi}=17.9 \pm 0.3 \pm 1.9 $, is experimentally known~\cite{Anastassov:2001cw} and related to $\hat g$. Similarly, one 
may also extract $\hat g$ from the unphysical decay $B^*\to B\pi$ in the chiral limit and where $m_b/\Lambda_\mathrm{QCD}$ corrections are better
controlled.

The coupling $g_{D^*\!D\pi}$ is related to a heavy-to-heavy transition form factor via the LSZ reduction of the pion and the use of PCAC,
$\pi(x) =  \partial^\mu A_\mu(x)/(f_\pi m_\pi^2)$, which leads to:
\begin{eqnarray}
   \lefteqn{  \langle  D(p_2) \pi( q) | D^*(p_1) \rangle  \  = \ q^\mu \frac{(m_\pi^2-q^2)}{f_\pi m_\pi^2}}   \nonumber \\
   & \times &  \!\!\!   \int d^4x\,  e^{i q\cdot x}   \langle  D(p_2) | A_\mu(x)  | D^*(p_1) \rangle .
\end{eqnarray}
Hence, the matrix element in Eq.~(\ref{H*Hpi}) has been reduced to the Fourier transform of a  transition matrix element between the $D^*$ and $D$ mesons
in the chiral limit with the axial QCD current $A_\mu(x) = \bar q^a \gamma_\mu\gamma_5 q^b$.  Results from this reduction procedure have been obtained 
on the lattice~\cite{deDivitiis:1998kj,Abada:2002xe} and most recently in a simulation with $n_f=2$~\cite{Becirevic:2009xp} which yields 
$g_{D^*\!D\pi}=20\pm 2$.

This form factor can also be calculated straightforwardly without reduction of the pion employing Eq.~(\ref{heavylightamp}) with the dressed quark 
propagators in Eq.~(\ref{sigmaSV}) and substituting the pion's BA~(\ref{BSApion}) for $\Gamma_I$. In a reassessment and improvement
of a calculation of $g_{D^*\!D\pi}$ within a Dyson-Schwinger model~\cite{Ivanov:1998ms}, we obtain $g_{D^*\!D\pi}=21$~\cite{El-Bennich2009}
in agreement with the lattice result~\cite{Becirevic:2009xp} and about $16\%$ larger than the experimental value.
\vspace*{-1mm}

\section{Conclusive remarks}

\vspace*{-1mm}
We have stressed the importance of hadronic effects in decays of heavy-flavoured mesons and portrayed the various theoretical {\em Ans\"atze} for 
the heavy-to-light transition form factors. In short, the main obstacle to their {\em precise\/} calculation, which veraciously reproduces the infrared features 
of QCD, are the uncertainties of the light-cone DA in the case of LCSR and the lack of  model-independent wave functions in relativistic quark model 
calculations. We have argued that the running quark mass of the DSE quark propagators is crucial to include confinement and DCSB effects in the
transition amplitudes; an unfinished task are consistent solutions of the BSE for the $D$ and $B$ mesons within the DSE formalism, which will reduce 
model dependence.

\vspace{-1mm}
\section*{Acknowledgments}

\vspace{-1mm}
Based on the talk given at Light Cone 2009: Relativistic Hadronic and Particle Physics, 8--13 July 2009, S\~ao Jos\'e dos Campos, S\~ao Paulo, Brazil.
B.~E. thanks the organisers at the Instituto Tecnol\'ogico de Aeron\'autica for the welcoming atmosphere and in particular Tobias Frederico and Jo\~ao 
Pacheco de Melo for their hospitality. Several stimulating discussions with Arlene Aguilar, Adriano Natale, Fernando Navarra, Marina Nielsen, 
Joannis Papavassiliou and Lauro Tomio were greatly appreciated. This work was supported by the Department of Energy, Office of Nuclear Physics, 
No. DE-AC02-06CH11357.

\end{document}